*Review Article*

# *Vitreoscilla* sp. hemoglobin (VHb) Expression in Heterologous Hosts: Some Aspects of Their Biotechnological Application

Haitham Qaralleh[b], Muhamad O. Al-limoun[a], Khaled M. Khleifat[a], Khaled Y Alsharafa[a], Amjad A. Al Tarawneh[c]

[a]Biology Department, Mutah University, Mutah, Karak, 61710, Jordan; [b]Department of Medical Laboratory Sciences, Mutah University, Mutah, Karak, 61710, Jordan; [c]Prince Faisal Center for Dead Sea, Environmental and Energy Research, Mutah University, Karak, Jordan

*Corresponding author: haitham@mutah.edu.jo

**Abstract:** It is worth mentioning that the high output of different physiological responses under the expression of vgb, may have a considerable effect on the enzyme productivity, dairy industry, heavy metal uptake, biodegradation of different organic pollutants and other applications. The expression of bacterial haemoglobin is useful in lessening the load of perceived toxic conditions such as high oxygen levels. This in turn probably has the same impact on some peripheral toxic materials. This, hemoglobin biotechnology can be extended to enhance production of pollutants degrading enzymes or production of some valuable manufacturing materials on the case-by-case bases. It is likely that the mechanism of bacterial hemoglobin (VHb) effects is intermediated via an oxygen trapping action. This may drive the enrichment of ATP production, which is mostly required for higher productivity of needed substances for that activity.

**Introduction**

*Vitreoscilla sp.* are strictly aerobic Gram-negative bacteria that live in oxygen-limited environments such as deep pools and cow dung (Pringsheim, 1951). *Vitreoscilla* hemoglobin (VtHb) was identified as hemoglobin by amino acid sequencing (Wakabayashi et al., 1986). Further studies showed that VtHb is composed of two identical subunits each containing a protoheme IX molecule; each subunit has a molecular weight of 16,392 Daltons.

In *Vitreoscilla* or *E. coli* transformed with *vgb*, VtHb levels were found to be increased when the cells were grown at low oxygen concentrations (Boerman and Webster, 1982; Dikshit et al., 1990). Further studies demonstrated that VtHb-containing *E. coli* cells grow more quickly and to a higher density compared with non-VtHb containing cells (Khosla et al., 1990; Khosravi et al., 1990; Joshi and Dikshit, 1994). It is thought that VtHb binds oxygen (particularly at low oxygen concentrations) and feeds it to the terminal oxidases of the respiratory chain (Wakabayashi et al., 1986). This theory was proposed based on the evidence mentioned above, for example, the fact that *Vitreoscilla* cells grown under oxygen limited conditions exhibit an approximately 50 fold increase in hem content (Boerman and Webster, 1982). Additionally, nearly half of the VtHb in *vgb*-bearing *E. coli* is located within the periplasmic space (Khosla et al., 1990), an ideal location for its proposed function. *Vgb* was also successfully cloned and expressed in members of the Pseudomonadaceae (Liu et al., 1995; Dikshit et al., 1990). In some cases, the presence of *vgb* enhanced the growth and metabolism of benzoic acid by members of this group (Liu et al., 1995; Kallio and Bailey, 1996). Degradation of nonpolar aromatic compounds usually requires the addition of oxygen at one or more steps (Suen et al., 1996). The presence of *vgb*/VtHb may provide oxygen directly to those oxygenases or stimulate benzoic acid metabolism by enhancing growth.

The presence of hemoglobin protein in prokaryotic organisms was first recognized in the gram-negative bacterium *Vitreoscilla* (Wakabayashi et al., 1986). This *Vitreoscilla* hemoglobin is the most widely studied bacterial hemoglobin, including its potential use in biotechnological applications (Khleifat and Abboud, 2003). Enhanced biosynthesis of VHb is mediated at the transcriptional level by an oxygen-sensitive promoter which is turned on under hypoxic conditions in the native and recombinant host of *E. coli* (Ramandeep et al., 2001; Joshi and Dikshit, 1994; Kallio et al., 1994).The expression of VHb hemoglobin may result in the enhancement of cell density, oxidative metabolism, bioremediation and engineered product formation especially under oxygen-limited conditions. A particular application of VHb expression involved the increased production of α-amylase (Khleifat and Abboud, 2003), the enhanced degradation of toxic wastes such as 2,4- DNT and benzoic acid degradation by Pseudomonads (Nasr et al., 2001). It is believed that the putative function of VHb is to trap oxygen and feeds it to a membrane terminal oxidase (Hwang et al., 2001). Presence of Vitreoscilla hemoglobin gene (vgb) on the plasmids minimized the negative impact conferred by such plasmid (Khosla and Bailey, 1989; Dikshit et al., 1992). Initial works on the VHb expressing gene, vgb and promoter activity in E. coli confirmed that the expression of the vgb gene is regulated by oxygen (Dikshit et al., 1989). It has been, also, suggested that besides its effective oxygen deliverity potential and transport function, VHb has generated fundamentally for the detoxification purpose (Minning et al., 1999). However, it has been hypothesized that VHb has the ability to trap oxygen molecules from neighboring domain and feeds it to the cellular activities in heterologous hosts due to its exceptional kinetic parameters for oxygen binding and freeing ($Kd = 72$ μM) (Webster, 1988; Frey et al., 2000; Kallio et al., 1994; Tsai et al., 1995; Akbas, 1997; Patel et al., 2000; Abboud et al., 2010; Khleifat et al., 2019a).

The general effect of VHb is based on the fact that a bacterial cells engineered with the gene (*vgb*) encoding *Vitreoscilla* hemoglobin (VHb) typically produce more protein than *vgb*-lacking cells (Khosla et al., 1990; Khleifat and Abboud, 2003)

**Tentative purification of VHb**
The purification data of *Vitreoscilla* hemoglobin from a culture of *E. coli* JM109 [pUC8:16] grown in terrific broth are shown in Table 1. The purified VtHb was examined by SDS-polyacrylamide gels stained with Coomassie blue R-250 (Fig. 1). Two bands of similar intensity were seen. Their molecular weights are 18,500 and 16,000 as determined by a standard curve of the same proteins (data not shown). Solutions of the purified VtHb were dark red, and the CO-difference spectrum obtained for the partially purified protein was used to calculate the VtHb concentration (data not shown). The ratio of $A_{410}/A_{280}$ after Sephadex G-100 was almost 1.5 which indicated that the use of Sephadex G-l00 as a single step after ammonium sulfate fractionation might be enough for partially purified protein. The additional Sephadex G-75 step produced a preparation with a $A_{410}/A_{280}$ ratio of almost 2.0, indicating a highly purified VHb preparation (data not shown), although this conclusion was clouded by the presence of two bands in these fractions.

Table 1. Purification Table Of *Vitreoscilla* Hemoglobin (Khleifat, 2010)

| Processes | Total $A_{410}$ | Total $A_{280}$ | Ratio $A_{410} : A_{280}$ | Yeild (%) | Purification (Fold) |
|---|---|---|---|---|---|
| **Crude Extract** | 1800 | 40450 | 0.044 | 100 | 1 |
| 45% Ammonium Sulphate Supernatant | 1599 | 24980 | .0.064 | 89 | 1.54 |
| 70% Amonium Sulphate Precipitate | 1200 | 9540 | 0.126 | 67 | 2.86 |
| Sephadex G-100 | 769.5 | 524.39 | 1.47 | 42 | 33.4 |
| Sephadex G-75 | 350.1 | 175.4 | 1.996 | 19 | 45.4 |

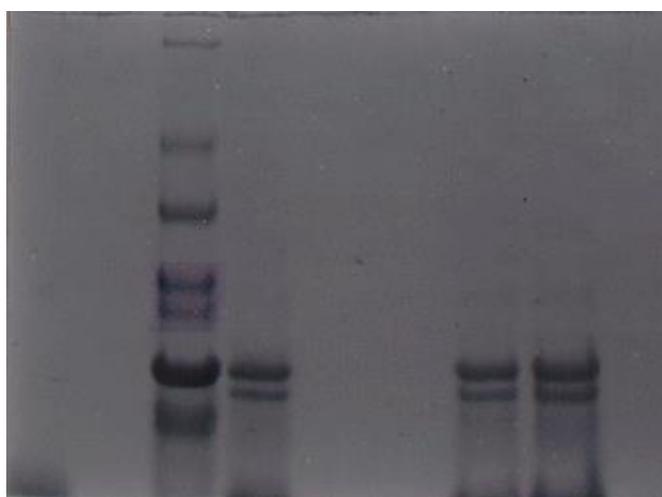

**Figure 1.** Purified VtHb by SDS-polyacrylamide gels stained with Coomassie blue R-250

**Does highly purified Vitreoscilla hemoglobin would in vitro affect partially purified DN'T dioxygenase**

Activity was tested at different concentrations of VtHb. The assays were run at both 100% and 25% air saturation (250 and 62.5 μM oxygen, respectively). VtHb inhibited DNT dioxygenase activity in a dose dependent manner, almost abolishing it at 6 μM (350 μg/ml) VtHb; the results are similar at both levels of oxygen. DNT dioxygenase was completely inhibited (at both 100% and 25% air saturation) by 25, 100 and 150 μg/ml horse hemoglobin and 25, 100 and 150 μg/ml horse myoglobin. In another independent experiment, DNT dioxygenase was shown to be as active in the presence of boiled VtHb compared with that VtHb absence, while the same amount of active VtHb completely inhibited the enzyme. The $K_d$ 's for both VHb and dioxygenase can be described by $K_d$ and $K_d'$, respectively, below. E indicates dioxygenase not bound to $O_2$ and E-$O_2$ represents the dioxygenase-$O_2$ complex (Khleifat, 2010).

$$Kd = \frac{(VHb)(O_2)}{(VHb-O_2)}$$

$$Kd' = \frac{(E)(O_2)}{(E-O_2)}$$

Solving the right hand equation for $O_2$ and substituting into the left hand equation gives:

$$\frac{(E-O_2)}{(E)} = \frac{(VHb-O_2)}{(VHb)} \cdot \frac{Kd}{Kd'}$$

At $O_2$ saturation (240 μM) and a typical VHb concentration (21 μM)

$$\frac{E-O_2}{E} = 4.9$$

without VHb and 4.5 with VHb (only an 8% decrease expected). Only at relatively high [VHb] and low [$O_2$] would one expect inhibition because the VHb would lower $O_2$ concentration and reduce

$$\frac{E-O_2}{E}$$

which is presumably the active form of the enzyme. Previous studies have shown that *vgb*/VtHb enhances the production of numerous products in bacteria and fungi. This includes the antibiotic actinorhodin, the synthesis of which requires molecular oxygen (De Modena et al., 1993). VtHb has also been linked to enhanced degradation of benzoate by *Xanthomonas* (Liu et al., 1996) and 2,4-DNT by *Burkholderia* (Patel et al, 2000) and *E. coli* bearing the DNT dioxygenase gene (Fish et al., submitted). All three processes require molecular oxygen. In the latter case, which was the precursor to the work reported here, the stimulation occurred during growth or when whole cells or whole cell extracts were tested and was probably due simply to an increase in DNT dioxygenase synthesis due to a VtHb produced increase in ATP levels in vivo (Tsai et al., 1995). Because the direct interaction of VtHb with DNT dioxygenase resulted in the inhibition of DNT conversion to MNC (Fig. 2), this interpretation, rather than a direct stimulation of DNT dioxygenase by VtHb, is supported. This could be a unique tool causes induction of the degradation ability of high concentrations of organic pollutants, having shorter times than those occurring in other micro-organisms (Khleifat, 2006a, Khleifat, 2006b; Shawabkeh et al., 2015; Khleifat, 2010; Khleifat, 2006c; Khleifat, 2007; Khleifat et al., 2015; Khleifat et al., 2019c)

A possible explanation for the apparent contradiction between the results of Fish and those presented here is that in recombinant *E. coli* VtHb results in greatly enhanced DNT dioxygenase production, which more than overcomes the direct inhibition of the enzyme by VtHb. This is consistent with the 3-4 fold increase in Vmax, indicating more enzyme, and the 2 fold increase in $K_M$ for DNT, indicating inhibition, in cells and cell extracts that also contained VTHb. The cause of this inhibition is most probably because of the competition between the two proteins (DNT dioxygenase and VtHb) for $O_2$. VtHb is likely to be effective in this competition because of its significantly higher oxygen affinity (compared with that of DNT dioxygenase). This could extended to the induction of surfactants biodegradation (Khleifat et al., 2008).

**Effect of Bacterial Hemoglobin on the Whey Lactose biodegradation**

Out of several bacterial strains being tested the recombinant *E. coli* strain that harbor the *vgb* gene proved to be the best bacteria for optimum whey lactose degradation and lactic acid production. It was superior than *Enterobacter* (WT) (Khleifat et al., 2006) as well as *E. coli* (WT) (Khleifat et al., 2006a; Tarawneh et al., 2009) and plasmid carrier *E. coli* (pUC9) that lacks *vgb* gene. Having selected the appropriate bacterial strain a good correlation was observed between bacterial biomass and lactic acid production. Hence, optimum conditions for bacterial growth such as temperature, pH, aeration, and yeast extract supplementation were also favored the lactic acid production (Abboud et al., 2010). Previous studies have shown that the genetic engineering strains of *E. coli* possess several advantages as biocatalysts for the production of pure lactic acid isomer (Patel *et al.*, 2000 and Dien *et al.*, 2001). The *E. coli* Lac Zy operon coding for B-galactosidase and lactose permease was inserted into xanthomonas campestris. The recombinant strain expressed high levels of B-galactosidase and grew well in a medium containing lactose as sole carbon source (Koffas et.al 1999). Therefore, it is likely that the superior production of lactic acid from whey lactose observed with *vgb* expression *E. coli* strain over the WT and pUC9 strains can be attributed to the improvement in bacterial mass production because of VHb protein induction (data not shown). Furthermore, it is known that the expression of *vgb* gene is VHb recombinant bacteria is capable of enhancing bacterial growth (Frey *et al*., 2002, Khosla and Bailey 1988, Khleifat *et al,* 2006, Khleifat and Abboud 2003).

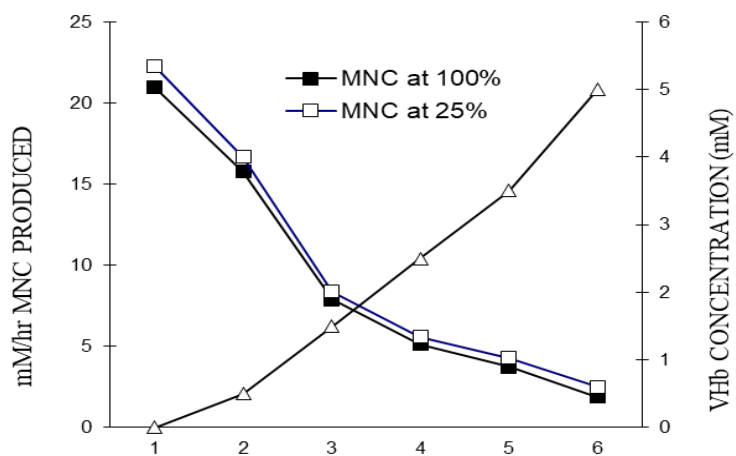

Figure 2. Effect of Vitreoscilla hemoglobin (VHb) on DNT dioxygenase activity. Experiments were made at hypoxic (25%) and hyperoxic (100%) conditions (Khleifat, 2010).

**Effect on Heavy metal Uptake**
When cadmium was added in vivo to different *Enterobacter* strains, it became toxic particularly at high concentrations (Fig. 3). This Toxicity effect of $Cd^{+2}$ was expressed by the rate of bacterial growth suppression relative to a control of bacterial growth without cadmium. The *vgb*-bearing cells persistently gave a better growth rate than the other two *vgb*-lacking control strains, when they were subjected to various concentrations of cadmium. The important role played by bacterial hemoglobin in metabolic activities (Kallio et al., 1994) makes this protein a suitable candidate for controlling heavy metal detoxification. It's known capability to enhance the production of cellular proteins (Kallio and Bailey, 1996) may either contribute in a general or selective manner to the $Cd^{+2}$ transporter protein activity.

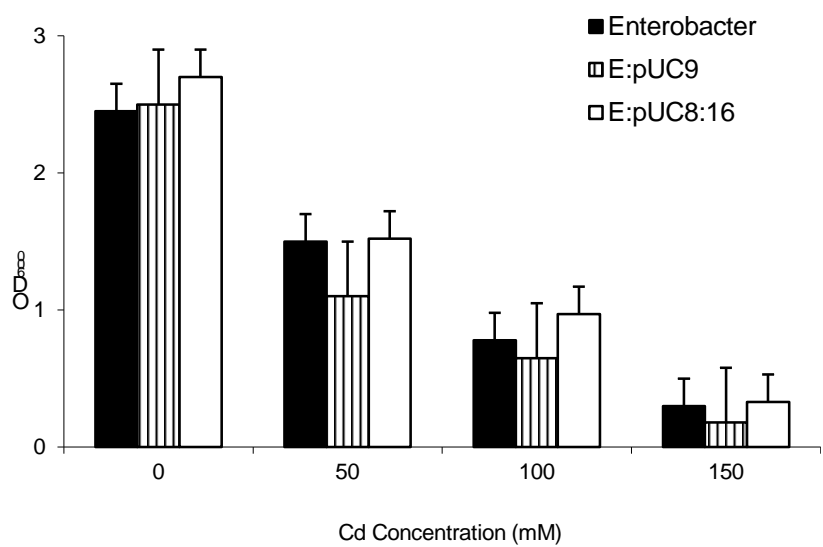

Figure 3 Toxicity of $Cd^{+2}$ against *Enterobacter aerogenes* strains (parental, E:pUC9 and E: pUC8:16) cultured in the LB buffered medium. $Cd^{+2}$ concentration used were 0, 50, 100 and 150 ppm. The growth curves based on $OD_{600}$ measurements were made, then the average of the first five $OD_{600}$ measurements of the stationary phase was taken as a final reading after 24 hours. Results represent the mean ± SD (n=3) (Khleifat et al., 2006b).

Table 2. Aeration dependence of Cd uptake, adsorption and biosorption by *Enterobacter aerogenes* (parental, E: pUC9 and E:pUC8:16) strains and their cell biomass. Values are the average of three individual experiments, standard deviations are in parenthesis (Khleifat et al., 2006b).

| Strain | Aeration (rev/min) | Adsorption Cd ppm/g biomass | Uptake Cd ppm/g biomass | Biosorption Cd ppm/g biomass | Cell biomass (g/L) |
|---|---|---|---|---|---|
| Enterobacter | 75 | 1.7 (0.2) | 15 (2.5) | 16.7 (3.2) | 3.40 (0.2) |
|  | 250 | 5.0 (0.3) | 37 (5.6) | 42 (7.1) | 5.95 (0.24) |
| *Enterobacter*:pUC9 | 75 | 1.3 (0.1) | 13.2 (3.1) | 14.5 (1.5) | 2.80 (0.15) |
|  | 250 | 6.0 (0.2) | 34.2 (6.5) | 40.2 (5.2) | 5.75 (0.42) |
| *Enterobacter*:pUC8:16 | 75 | 0.7 (0.05) | 35.7 (4.5) | 36.4 (4.2) | 5.10 (0.3) |
|  | 250 | Nd | 56.2 (8.7) | 56.2 (8.3) | 7.20 (0.46) |

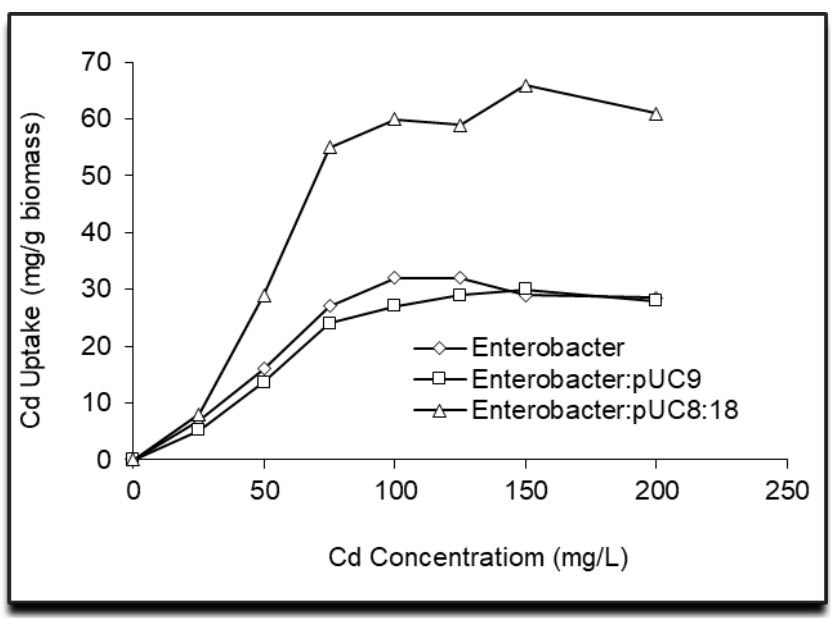

Figure 4. Effect of $Cd^{+2}$ concentration in the medium on $Cd^{+2}$ uptake by *Enterobacter aerogenes* strains (parental, E:pUC9 and E: pUC8:16). The uptake assay was performed at 37 °C as described in materials and methods. Each point represents the mean ± SD (n=3) (Khleifat et al., 2006b)

The *vgb*-harboring strains showed better uptake of cadmium than *vgb*-lacking strains (Fig 4 and Table 2). Under low aeration there was two fold enhancement of $Cd^{+2}$ uptake in *vgb*-harboring strains compared with 1.6 fold enhancement under high aeration (Table 2). The important role played by bacterial hemoglobin in metabolic activities (Kallio et al., 1994) makes this protein a suitable candidate for controlling heavy metal detoxification. VHb Expression via heterologous hosts often improves growth, as well as rising the yields of recombinant and total proteins and enhances the production of antibiotics and α-amylase especially under hypoxic status and hyperoxic conditions (Dikshit et al., 1990; Khosla and Bailey, 1988; Khosravi et al., 1990; Kallio and Bailey, 1996; Park et al., 2002; Khleifat and Abboud, 2003). It may either contribute in a general or selective manner to the $Cd^{+2}$ transporter protein activity. The enhancement in $Cd^{+2}$ uptake by the VHb bearing strain under low aeration is in keeping with the function of VHb as an oxygen-trap protein. The oxy-VHb may act positively on some key redox component of the cell which could be a sensor, a regulator or an allosteric site of respiratory protein (Kallio et al., 1994; Liu et al., 1995; Nies and Silver, 1995; Ramandeep et al., 2001). This may explain the high sensitivity of $Cd^{+2}$ uptake to the inhibition of uncouplers CCCP and sodium azide observed particularly in the *vgb*-bearing transformed strain. The slight inhibition of $Cd^{+2}$ uptake by thiol compounds might be the outcome of an extra cellular complexation between the $Cd^{+2}$ and the thiol compound. A Cd-S complex has been demonstrated in *Klebsiella pneumoniae* and *Saccharomyces cerevisiae* as a mechanism of $Cd^{+2}$ detoxification (Aiking et al., 1985; Nies, 1995; Gharieb and Gadd, 2004)

In conclusion, evidence are presented in this work for an active transport system of $Cd^{+2}$ in *Enterobacter aerogenes*. This evidence is indicated by the Michaelis–Menten saturation pattern of $Cd^{+2}$ uptake, the decrease of this transport rate by inhibitors of cellular energy and thiol compounds which was then particularly relieved by VHb expression (Scott and Palmer, 1990). The impact of this improvement on bacterial bioremediation of toxic heavy metals will await further research work.

Copper uptake by *E. coli* (wild type), *E. coli*: pUC8:16 and *E. coli*: pUC9 pre-grown on LB media: It's was shown that cells grown on LB broth resulted in lower level of copper uptake than obtained with minimal media irrespective of carbon and nitrogen source used. It was reported that cells grown on minimal media showed significant higher expression rate of many genes involved in biosynthesis of building blocks, particularly the amino acid biosynthetic pathways (Tao *et al.*, 1999; Khleifat, 2006). The increase order was *E. coli*: pUC8: 16<*E. coli*: pUC9<*E. coli* (wild type) with maximum $Cu^{+2}$ uptake amount of 811, 523 and 440 ppm $g^{-1}$ biomass, respectively. These maximum values were determined at initial $CuSO_4$ concentration of 100 ppm (Khleifat and Abboud, 2003; Khleifat et al., 2006a).

We have shown that the VHb-expressing *E. coli*, which bears *Vitreoscilla* hemoglobin gene, *vgb*, is capable of improving the lead biosorption (Dikshit and Webster, 1988; Khleifat, 2006d; Khleifat et al., 2006b; Tarawneh et al., 2009). To the contrary, control strains of parental cells have failed to achieve a similar improvement in Pb sorption. Furthermore, kinetic data obtained from the effect of Pb on bacterial growth suggest that VHb-expressing *E. coli* strains endure better resistance to Pb toxicity than the control strain. Other considerations is possible to examine bacterial hemoglobin and its relationship to the antioxidant and antimicrobial activities of some plants extracts and their oils (Khleifat & Homady, 2000; Khleifat et al., 2006a; Khleifat et al., 2008; Abboud,et al., 2009; Khleifat et al., 2010; Zeidan et al., 2013; Althunibat et al., 2010; Khleifat et al., 2014; Majali et al., 2015; Althunibat et al., 2016; Al-Asoufi et al., 2017; Khleifat et al., 2019b; Qaralleh et al., 2019; Al-Limoun et al., 2019). The mechanism of bacterial hemoglobin (VHb) effects is likely to be mediated by oxygen trapping. This may enrich the production of ATP, which is often required to increase the productivity of the materials required for this activity.

## References

Abboud, M. M., Aljundi, I. H., Khleifat, K. M., & Dmour, S. (2010). Biodegradation kinetics and modeling of whey lactose by bacterial hemoglobin VHb-expressing Escherichia coli strain. *Biochemical Engineering Journal*, *48*(2), 166-172.

Abboud, M. M., Saeed, H. A., Tarawneh, K. A., Khleifat, K. M., & Al Tarawneh, A. (2009). Copper uptake by Pseudomonas aeruginosa isolated from infected burn patients. *Current microbiology*, *59*(3), 282-287.

Aiking, H., Govers, H., & Van't Riet, J. (1985). Detoxification of mercury, cadmium, and lead in Klebsiella aerogenes NCTC 418 growing in continuous culture. *Appl. Environ. Microbiol.*, *50*(5), 1262-1267.

Al-Limoun, M. O., Khleifat, K. M., Alsharafa, K. Y., Qaralleh, H. N., & Alrawashdeh, S. A. (2019). Purification and characterization of a mesophilic organic solvent tolerant lipase produced by Acinetobacter sp. K5b4. *Biocatalysis and Biotransformation*, *37*(2), 139-151.

Althunibat, O. Y., Qaralleh, H., Al-Dalin, S. Y. A., Abboud, M., Khleifat, K., Majali, I. S., ... & Jaafraa, A. (2016). Effect of thymol and carvacrol, the major components of Thymus capitatus on the growth of Pseudomonas aeruginosa. *J Pure Appl Microbiol*, *10*, 367-74.

Althunibat, O. Y., Al-Mustafa, A. H., Tarawneh, K., Khleifat, K. M., Ridzwan, B. H., & Qaralleh, H. N. (2010). Protective role of Punica granatum L. peel extract against oxidative damage in experimental diabetic rats. *Process Biochemistry*, *45*(4), 581-585.

Al-Asoufi, A., Khlaifat, A., Tarawneh, A., Alsharafa, K., Al-Limoun, M., & Khleifat, K. (2017). Bacterial quality of urinary tract infections in diabetic and non-diabetics of the population of Ma'an Province, Jordan. *Pakistan J Biol Sci*, *20*, 179-88.

Boerman, S. J., & Webester, D. A. (1982). Control of heme content in Vitreoscilla by oxygen. *The Journal of General and Applied Microbiology*, *28*(1), 35-43.

Dikshit, K. L., Dikshit, R. P., & Webster, D. A. (1990). Study of Vitreoscilla globin (vgb) gene expression and promoter activity in E. coli through transcriptional fusion. *Nucleic Acids Research*, *18*(14), 4149-4155.

Dikshit, K. L., & Webster, D. A. (1988). Cloning, characterization and expression of the bacterial globin gene from Vitreoscilla in Escherichia coli. *Gene*, *70*(2), 377-386.

Dikshit, R. P., Dikshit, K. L., Liu, Y., & Webster, D. A. (1992). The bacterial hemoglobin from Vitreoscilla can support the aerobic growth of Escherichia coli lacking terminal oxidases. *Archives of Biochemistry and Biophysics*, *293*(2), 241-245.

Dikshit, K. L., Spaulding, D., Braun, A., & Webster, D. A. (1989). Oxygen inhibition of globin gene transcription and bacterial haemoglobin synthesis in Vitreoscilla. Microbiology, 135(10), 2601-2609.

Frey, A. D., Farrés, J., Bollinger, C. J., & Kallio, P. T. (2002). Bacterial hemoglobins and flavohemoglobins for alleviation of nitrosative stress in Escherichia coli. *Appl. Environ. Microbiol.*, *68*(10), 4835-4840.

Frey, A.D., J.E. Bailey and P.T. Kallio, 2000. Expression of Alcaligenes Eutrophus and engineered Vitreoscilla hemoglobin-reductase fusion protein for improved hypoxic growth of Escherichia coli. Applied Environ. Microbiol., 66: 98-104.

Gharieb, M. M., & Gadd, G. M. (2004). Role of glutathione in detoxification of metal (loid) s by Saccharomyces cerevisiae. *Biometals*, *17*(2), 183-188.

Joshi, M., & Dikshit, K. L. (1994). Oxygen-dependent regulation of Vitreoscilla globin gene: evidence for positive regulation by FNR. *Biochemical and biophysical research communications*, *202*(1), 535-542.

Kallio, P. T., Bailey, J. E. 1996. Expression of vhb encoding Vitreoscillu hemoglobin (VHb) enhances total protein secretion and improves the production of a-amylase and neutral protease in Bacillus subtilis. Biotechnol. Prog. 12: 31-39.

Kallio PT, Kim DJ, Tsai PS, et al. (1994) Intracellular expression of Vitreoscilla hemoglobin alters Escherichia coli energy metabolism under oxygen-limited conditions. Eur J Biochem 15(1–2): 201–208.

Khazaeli, M. B., & Mitra, R. S. (1981). Cadmium-binding component in Escherichia coli during accommodation to low levels of this ion. *Appl. Environ. Microbiol.*, *41*(1), 46-50.

Khleifat, K. M. (2010). Characterization of 2, 4-Dinitrotoluene Dioxygenase from Recombinant Esherichia coli Strain PFJS39: Its Direct Interaction with Vitreoscilla Hemoglobin. *Bioremediation Journal*, *14*(1), 38-53.

Khleifat, K. M. (2006a). Correlation Between Bacterial Hemoglobin and Carbon Sources: Their Effect on Copper Uptake by Transformed E. coli Strain αDH5. *Current microbiology*, *52*(1), 64-68.

Khleifat, K. M. (2010). Characterization of 2, 4-Dinitrotoluene Dioxygenase from Recombinant Esherichia coli Strain PFJS39: Its Direct Interaction with Vitreoscilla Hemoglobin. *Bioremediation Journal, 14*(1), 38-53.

Khleifat, K. M. (2006b). Biodegradation of phenol by Ewingella americana: Effect of carbon starvation and some growth conditions. *Process Biochemistry, 41*(9), 2010-2016.

Khleifat, K. M. (2006c). Biodegradation of sodium lauryl ether sulfate (SLES) by two different bacterial consortia. *Current microbiology, 53*(5), 444-448.

Khleifat, K. M. (2006d). Biodegradation of linear alkylbenzene sulfonate by a two-member facultative anaerobic bacterial consortium. *Enzyme and microbial technology, 39*(5), 1030-1035.

Khleifat, K. M. (2007). Effect of substrate adaptation, carbon starvation and cell density on the biodegradation of phenol by Actinobacillus sp. *Fresenius Environmental Bulletin, 16*(7), 726-730.

Khleifat, K. M., Sharaf, E. F., & Al-limoun, M. O. (2015). Biodegradation of 2-chlorobenzoic acid by enterobacter cloacae: Growth kinetics and effect of growth conditions. *Bioremediation Journal, 19*(3), 207-217.

Khleifat, K. M., Al-limoun, M. O., Alsharafa, K. Y., Qaralleh, H., & Al Tarawneh, A. A. (2019a). Tendency of using different aromatic compounds as substrates by 2, 4-DNT dioxygenase expressed by pJS39 carrying the gene dntA from Burkholderia sp. strain DNT. *Bioremediation Journal, 23*(1), 22-31.

Khleifat, K. M., Matar, S. A., Jaafreh, M., Qaralleh, H., Al-limoun, M. O., & Alsharafa, K. Y. (2019b). Essential Oil of Centaurea damascena Aerial Parts, Antibacterial and Synergistic Effect. *Journal of Essential Oil Bearing Plants*, *22*(2), 356-367.

Khleifat, K. M., Abboud, M., Laymun, M., Al-Sharafa, K., & Tarawneh, K. (2006a). Effect of variation in copper sources and growth conditions on the copper uptake by bacterial hemoglobin gene (vgb) bearing E. coli. *Pakistan Journal of Biological Sciences*, *9*(11), 2022-2031.

Khleifat, K., Abboud, M., Al-Shamayleh, W., Jiries, A., & Tarawneh, K. (2006c). Effect of chlorination treatment on gram negative bacterial composition of recycled wastewater. *Pak. J. Biol. Sci*, *9*, 1660-1668.

Khleifat, K., & Abboud, M. M. (2003). Correlation between bacterial haemoglobin gene (vgb) and aeration: their effect on the growth and α-amylase activity in transformed Enterobacter aerogenes. *Journal of applied microbiology*, *94*(6), 1052-1058.

Khleifat, K. M., Tarawneh, K. A., Wedyan, M. A., Al-Tarawneh, A. A., & Al Sharafa, K. (2008). Growth kinetics and toxicity of Enterobacter cloacae grown on linear alkylbenzene sulfonate as sole carbon source. *Current microbiology*, *57*(4), 364-370.

Khleifat, K. M., Hanafy, A. M. M., & Al Omari, J. (2014). Prevalence and molecular diversity of Legionella pneumophila in domestic hot water systems of private apartments. *British Microbiology Research Journal*, *4*(3), 306.

Khleifat, K. M., Abboud, M. M., & Al-Mustafa, A. H. (2006b). Effect of Vitreoscilla hemoglobin gene (vgb) and metabolic inhibitors on cadmium uptake by the heterologous host Enterobacter aerogenes. *Process Biochemistry*, *41*(4), 930-934.

Khleifat, K., & Homady, H. M. (2000). Bacterial hemoglobin gene (vgb) transformed into Escherichia coli enhances lead-uptake and minimizes it's adsorption. *Pakistan Journal of Biological Sciences*, *3*(9), 1480-1483.

Khleifat, K. M., Halasah, R. A., Tarawneh, K. A., Halasah, Z., Shawabkeh, R., & Wedyan, M. A. (2010). Biodegradation of linear alkylbenzene sulfonate by Burkholderia sp.: Effect of some growth conditions. *Int J Agr Biol*, *12*, 17-25.

Khleifat, K. M., Matar, S. A., Jaafreh, M., Qaralleh, H., Al-limoun, M. O., & Alsharafa, K. Y. (2019c). Essential Oil of *Centaurea damascena* Aerial Parts, Antibacterial and Synergistic Effect. *Journal of Essential Oil Bearing Plants*, *22*(2), 356-367.

Koffas, M., Roberge, C., Lee, K., & Stephanopoulos, G. (1999). Metabolic engineering. *Annual Review of Biomedical Engineering*, *1*(1), 535-557.

Khosla C, Bailey JE (1988) Heterologous expression of a bacterial hemoglobin improves the growth properties of recombinant E. coli. Nature 33:633–635

Khosla, C., & Bailey, J. E. (1989). Characterization of the oxygen-dependent promoter of the Vitreoscilla hemoglobin gene in Escherichia coli. Journal of bacteriology, 171(11), 5995-6004.

Khosravi, M., Webster, D. A., & Stark, B. C. (1990). Presence of the bacterial hemoglobin gene improves α-amylase production of a recombinantEscherichia coli strain. *Plasmid*, *24*(3), 190-194.

Liu, S.C., Webster, D.A. and Stark, B.C. (1995). Cloning and expression of the *Vitreoscilla* hemoglobin gene in *Pseudomonas*: effects on cell growth. *Applied Microbiology and Biotechnology*, 44 419-424.

Majali, I. S., Oran, S. A., Khaled, M. K., Qaralleh, H., Rayyan, W. A., & Althunibat, O. Y. (2015). Assessment of the antibacterial effects of Moringa peregrina extracts. *African Journal of microbiology research*, *9*(51), 2410-2414.

Minning, D.M., A.J. Gow, J. Bonavenuras, R. Braun, M. Dewhirst, D.E. Goldberg and J.S. Stamler, 1999. Ascaris hemoglobin is a nitric oxide-activated deoxygenase. Nature, 401: 497-502.

Nasr, M. A., Hwang, K. W., Akbas, M., Webster, D. A., & Stark, B. C. (2001). Effects of Culture Conditions on Enhancement of 2, 4-Dinitrotoluene Degradation by BurkholderiaEngineered with the Vitreoscilla Hemoglobin Gene. *Biotechnology progress*, *17*(2), 359-361.

Nies, D. H. (1995). The cobalt, zinc, and cadmium efflux system CzcABC from Alcaligenes eutrophus functions as a cation-proton antiporter in Escherichia coli. *Journal of bacteriology*, *177*(10), 2707-2712.

Nies, D. H., & Silver, S. (1995). Ion efflux systems involved in bacterial metal resistances. *Journal of industrial microbiology*, *14*(2), 186-199.

Patel, S. M., Stark, B. C., Hwang, K. W., Dikshit, K. L., & Webster, D. A. (2000). Cloning and expression of Vitreoscilla hemoglobin gene in Burkholderia sp. strain DNT for enhancement of 2, 4-dinitrotoluene degradation. *Biotechnology progress*, *16*(1), 26-30.

Park, K. W., Kim, K. J., Howard, A. J., Stark, B. C., & Webster, D. A. (2002). Vitreoscilla hemoglobin binds to subunit I of cytochrome bo ubiquinol oxidases. *Journal of Biological Chemistry*, *277*(36), 33334-33337.

Qaralleh, H., Khleifat, K. M., Al-Limoun, M. O., Alzedaneen, F. Y., & Al-Tawarah, N. (2019). Antibacterial and synergistic effect of biosynthesized silver nanoparticles using the fungi Tritirachium oryzae W5H with essential oil of Centaurea damascena to enhance conventional antibiotics activity. *Advances in Natural Sciences: Nanoscience and Nanotechnology*, *10*(2), 025016.

Ramandeep, Hwang KW, Raje M, et al. (2001) Vitreoscilla hemoglobin. Intracellular localization and binding to membranes. J Biol Chem 276(27):2481–249.

Scott, J. A., & Palmer, S. J. (1990). Sites of cadmiun uptake in bacteria used for biosorption. *Applied Microbiology and Biotechnology*, *33*(2), 221-225.

Shawabkeh, R., Khleifat, K. M., Al-Majali, I., & Tarawneh, K. (2007). Rate of biodegradation of phenol by Klebsiella oxytoca in minimal medium and nutrient broth conditions. *Bioremediation Journal, 11*(1), 13-19.

Suen, W. C., Haigler, B. E., & Spain, J. C. (1996). 2, 4-Dinitrotoluene dioxygenase from Burkholderia sp. strain DNT: similarity to naphthalene dioxygenase. *Journal of bacteriology*, *178*(16), 4926-4934.

Tarawneh, K. A., Al-Tawarah, N. M., Abdel-Ghani, A. H., Al-Majali, A. M., & Khleifat, K. M. (2009). Characterization of verotoxigenic Escherichia coli (VTEC) isolates from faeces of small ruminants and environmental samples in Southern Jordan. *Journal of basic microbiology*, *49*(3), 310-317.

Tao, H., Bausch, C., Richmond, C., Blattner, F. R., & Conway, T. (1999). Functional genomics: expression analysis ofEscherichia coli growing on minimal and rich media. *Journal of bacteriology*, *181*(20), 6425-6440.

Tsai, P. S., Rao, G., & Bailey, J. E. (1995). Improvement of Escherichia coli microaerobic oxygen metabolism by Vitreoscilla hemoglobin: new insights from NAD (P) H fluorescence and culture redox potential. Biotechnology and bioengineering, 47(3), 347-354.

Wakabayashi, S., Matsubara, H. and Webster, D.A., Primary sequence of a dimeric bacterial hemoglobin from *Vitreoscilla*. *Nature* (London), 1986, 322 481-483.

Webster, D. A. (1988). Structure and function of bacterial hemoglobin and related proteins. Advances in inorganic biochemistry, 7, 245-265.

Pringshem, E.G., The *Vitreoscillaceae*: A family of colorless, gliding, filamentous organisms . *Journal of General Microbiology*, 1951, 5 124-149.

Khosla, C., Curtis , J.E., DeModena , J.A., Rinas, U. and Bailey, J.E., Expression of intracellular hemoglobin improves protein synthesis in Oxygen-limited *Escherichia coli*. *Bio/technology*, 1990, 8 849-853.

Hwang, K. W., Raje, M., Kim, K. J., Stark, B. C., Dikshit, K. L., & Webster, D. A. (2001). Vitreoscilla Hemoglobin intracellular localization AND binding TO membranes. *Journal of Biological Chemistry*, *276*(27), 24781-24789.

Zeidan, R., Oran, S., Khleifat, K., & Matar, S. (2013). Antimicrobial activity of leaf and fruit extracts of Jordanian Rubus sanguineus Friv.(Rosaceae). *African Journal of Microbiology Research*, *7*(44), 5114-5118.